# Ultra-Sensitive Chip-Based Photonic Temperature Sensor Using Ring Resonator Structures

Haitan Xu[1], Mohammad Hafezi[1], J. Fan[1], J. M. Taylor[1,2] and G. F. Strouse[3], Zeeshan Ahmed[3]

[1]*Joint Quantum Institute, University of Maryland, College Park, MD 20742*

[2]*National Institute of Standards and Technology, Gaithersburg, MD 20899*

[3]*Thermodynamic Metrology Group, Sensor Science Division, Physical Measurement Laboratory, National Institute of Standards and Technology, Gaithersburg, MD 20899*

**ABSTRACT**

Resistance thermometry provides a time-tested method for taking temperature measurements. However, fundamental limits to resistance-based approaches has produced considerable interest in developing photonic temperature sensors to leverage advances in frequency metrology and to achieve greater mechanical and environmental stability. Here we show that silicon-based optical ring resonator devices can resolve temperature differences of 1 mK using the traditional wavelength scanning methodology. An even lower noise floor of 80 μK for measuring temperature difference is achieved in the side-of-fringe, constant power mode measurement.

Temperature measurements play a central role in modern life ranging from process control in manufacturing[1], physiological monitoring[2] and tissue ablation[3] in medicine, and environmental control and monitoring in buildings[4] and automobiles[5]. Despite the ubiquity of thermometers, the underlying technology has been slow to advance over the last century.[6] The standard bearer for accurate temperature measurement, the standard platinum resistance thermometer (SPRT) was initially developed over a century ago.[6,7] Furthermore, many modern temperature sensors still rely on resistance measurements of a thin metal film or wire whose resistance varies with temperature.[6] Though resistance thermometers can routinely measure temperature with uncertainties of 10 mK, they are sensitive to mechanical shock which causes the resistance to drift over time requiring frequent off-line, expensive, and time consuming calibrations.[7]

In recent years there has been considerable interest in developing photonic devices as an alternative to resistance thermometers[8-10] as they have the potential to provide greater temperature sensitivity while being robust against mechanical shock and electromagnetic interference. Furthermore, the low weight, small form factor photonic devices might be multiplexed to provide a low-cost sensing solutions.

Photonic temperature sensors exploit temperature dependent changes in a material's properties – typically, a combination of thermo-optic effect and thermal expansion.[11-13] For example, one of us (Strouse[14]) has demonstrated that the intrinsic temperature dependence of a synthetic sapphire's refractive index can be exploited for highly sensitive temperature measurement by measuring microwave frequency shifts of monocrystalline sapphire's resonant whispering gallery modes, with measurement uncertainties of 10 mK from 273.15 K to 373 K.[14] An optical analog of this, using infrared light to probe strain-free fiber Bragg gratings (FBG), exhibits temperature dependent shifts in resonant wavelength of 10 pm/K.[8,9,11,12] This sensitivity has been exploited in commercially available photonic temperature sensors.[8,11,12]

However, FBGs are susceptible to strain and are relatively large. Instead, we consider the use of ring resonators. In recent years, ring resonator[15] based devices have been exploited for bio-chemical sensing applications.[16] In these sensors, temperature induced shifts in resonance frequency have been a complicating variable or feature that can adversely impact sensor sensitivity and specificity. Indeed, development of an athermal ring resonator is an active area of research.[17] Here, we examine their use as thermal sensors, extending the concepts from a recent study Kim *et al*[18] that demonstrated that silicon ring resonator devices respond rapidly to small temperature variations (in about 6 μs).

The temperature dependence of the ring resonator arises from temperature-induced changes in refractive index (n) and in the physical dimensions of the ring. A qualitative analysis of a ring resonator yields a resonance wavelength for a single ring resonator of:

$$\lambda_m = [n_{eff}(\lambda_m, T) * L(T)] / m \qquad (1)$$

where $\lambda_m$ is the vacuum wavelength, $n_{eff}$ is the effective refractive index, m is the mode number, L is the ring perimeter, and T is the temperature. Thus, the temperature-induced shift in wavelength is given by:

$$\Delta\lambda_m = \left\{ \frac{\left[\left(\frac{\partial n}{\partial T}\right) + n\left(\frac{\partial L}{\partial T}\right)\left(\frac{1}{L}\right)\right]}{n_g} \right\} * (\Delta T * \lambda_m) \quad (2)$$

Where the group index is $n_g = [n - \lambda_m \left(\frac{\partial n}{\partial \lambda_m}\right)]$. The variation in the refractive index due to the thermal expansion coefficient for silicon (3.57 x $10^{-6}$/K) is a factor of 100 smaller than that of the estimated thermo-optic effect (2 x $10^{-4}$/K) of the silicon waveguide and thus not included in our analysis of the performance of ring resonator devices.

Fig. 1A depicts a schematic of the photonic device consisting of a ring resonator coupled to a straight-probe waveguide, with cross-section designed to be 610 nm × 220 nm to assure single-mode propagation of the transverse-electric (TE) light (the electric field in the slab plane) at the telecom wavelengths (1550 nm) and air gap of 130 nm for evanescent coupling between the resonators and the probing waveguide. The photonic chip was fabricated using standard CMOS techniques on a silicon on insulator (SOI) wafer with a 220 nm thick layer of silicon on top of a 2 μm thick buried oxide layer that isolates the optical mode and prevents loss to the substrate. The fabrication of silicon devices was performed through ePIXfab, and set-up by IMEC facility. The masks were made using deep ultraviolet 193 nm photolithography and were etched in two steps for fabricating waveguides and gratings, respectively. The grating provides a means for efficient free space coupling of light in/out of the photonic device (about 4 dB per coupler). The process was followed by thermal oxidation (10 nm) to reduce surface roughness.

In our experiments, a tunable Santec laser (TSL-210)[19] was used to probe the ring resonator (see Fig 1). A small amount of laser power was immediately picked up from the laser output for wavelength monitoring (HighFinesse WS/7)[19] while the rest, after passing through the photonic device, was detected by a large sensing-area power meter (Newport, model 1936-R[19]). The photonic chip itself is mounted on a 2-axis stage (Newport[19]) in a two-stage temperature controlled enclosure. Input from a platinum resistance thermometer from each stage is feed to its respective proportional-integral-derivative controller that drives a thermoelectric cooler (Laird Technologies[19]). The first stage minimizes global temperature fluctuations inside the enclosure to better than ± 1 K, while the second stage minimizes temperature variations at the chip to ≤ 17 mK over 24 hours. The relative humidity (%RH) levels inside the temperature enclosure were monitored using a portable humidity meter (Vaisala[19]). Relative humidity levels were varied inside the enclosure by varying the flow rate of water saturated air.

Traditionally, photonic thermometers such as those based on Fiber Bragg gratings (FBG) employ continuous wavelength scanning techniques to measure temperature changes.[12] In this scheme -- *Wavelength Scanning mode* -- the probe laser is continuously scanned across the frequency region of interest and the transmission/reflection spectrum is recorded and its center frequency computed. With the help of a previously determined calibration curve, the center frequency is then converted to temperature. Using this approach we evaluated our ring resonator based temperature sensor by systematically varying the temperature between 288 K and 306 K. Our measurements of a resonator with ring diameter of 11 μm and a gap of 130 nm, show a free spectral range (FSR) of ca. 9.2 nm and bandwidth of 0.03 nm corresponding to a quality factor (Q-factor) of 52000 (Fig 1C). Using COMSOL we estimate a group index of 4.2.  As shown in Fig. 1D, we observed that this device shows a temperature dependent shift in resonance wavelength of 77 pm/K, which is an eight times improvement over FBG temperature sensor. Given the wavelength resolution of 0.1 pm for the current setup, we estimate that this temperature sensor can resolve temperature differences of $\approx$ 1 mK.

Humidity and intrinsic heating are known to often affect the optical response of photonic sensors.[20] We systematically examined the impact of relative humidity (%RH) changes on resonance frequency by varying humidity levels from 17 %RH to 26 %RH at 294.15 K. Our measurements indicate humidity does not have any significant impact on the resonance frequency (data not shown). The insensitivity of our device to changes in humidity likely derives from the passivating $SiO_2$ layer deposited on-top of the ring resonator. In contrast to the observed insensitivity to humidity changes, we find that varying the input laser power from 0.0063 mW to 3.16 mW results in a systematic upshift in the resonance frequency and a corresponding systematic error in temperature determination (see Fig 2A). As shown in Fig 2A, as the incident laser power is increased, the ring resonator increasingly over-estimates the ambient temperature as compared to the platinum resistance thermometer located on the chip stage. Our results suggest that incident laser power fluctuations, if left unaccounted for may be a significant source of measurement uncertainty. The impact of laser induced self-heating, however, might be mitigated by operating at low laser powers (≤12 μW) or by ensuring the operating laser power matches the laser power used during device calibration of the sensor.

Here we demonstrate an alternative approach to *Wavelength Scanning mode* for measuring temperature changes smaller than 1 mK. In this measurement scheme -- *side of fringe, constant power mode* -- the laser power is maintained at a constant power level and its frequency centered on the side of resonance at the point of steepest descent (point of half-contrast). A small, temperature-dependent shift in resonance frequency results in a large change in transmitted laser intensity. Since the resonance lineshape is known, intensity fluctuations can be converted to an estimate of center frequency change without measuring the peak shape. This measurement scheme is limited by the system noise including both laser frequency and intensity noise, as shown in Fig. 2B. We have quantified the system noise by systematically measuring the output power of the ring resonator as a function of time with laser wavelength set on- and off-resonance (Table 1). The laser frequency was simultaneously monitored during these measurements to quantify laser frequency noise which primarily derives from long term drift in laser frequency. To quantify this, we performed a control, off-resonance measurement in which the laser is tuned in between two resonance peaks; transmitted intensity corresponds to light travelling through the entire optical train except the on-chip resonator. This control measurement allows us to estimate noise sources deriving from fluctuations in laser power and light coupling. The side-of-fridge approach is estimating the resonance frequency by local inversion of the expected lineshape, and thus contains additional contributions from background thermal fluctuations and laser frequency noise. With the data sampling rate set at 100 Hz, a total of 32,768 data points were collected for each measurement. This data set was used to generate the noise power spectra density spectrum[21] and the corresponding Allan variance plot[13] (Fig 2B and 2C). Our results indicate that Allan variance reaches a minimum at sampling rate of ≈1 second. At 296.15 K, the noise floor on the temperature measurement of ≈80 μK is set by on-resonance measurement, which represents a 13-fold improvement over the traditional wavelength scanning mode. The on-peak measurement contains contributions from laser power noise (≈ 10 μK), laser frequency noise (≈ 40 μK) and additional technical noise which we attribute to background thermal fluctuations (≈ 69 μK). At time scales of 1 min or longer (time required for scanning the resonance peak), the noise floor for side-of-fringe mode rises to ±5 K. Thus our results suggest that for measurement applications requiring long observation times (e.g. process control), in the absence of frequency stabilization, the classic wavelength scanning mode provides lower measurement uncertainty. The side of fringe constant power mode is best suited for dynamic temperature measurements.

In summary, we have demonstrated that ring resonators can be used for temperature sensing with sub-mK resolution. The noise level of these devices can be further reduced by laser frequency and power stabilization techniques. The low noise levels of these photonic sensors along with their low thermal mass and immunity to electro-magnetic interference make them attractive choice for applications in aerospace and microfluidics.  However, key systematic errors induced by laser power and laser intensity noise remain to be addressed over a wider range of temperature and operating environments.

**ACKNOWLEDGEMENTS**

We thank Kevin Douglass, Stephen Maxwell, Howard Yoon, Joe Hodges for helpful comments. This work was supported in part by DARPA QuASAR and by the NSF-funded Physics Frontier Center at the JQI.

**Table 1: Noise Sources Contributing to Photonic Temperature Measurement**

| *Noise Source* | *Noise Level (μK)* |
| --- | --- |
| Laser Power | 10 |
| Laser Frequency | 40 |
| Background thermal fluctuations | 69 |

## FIGURES

**Figure 1: A) SEM image of ring resonator device (11 µm radius, 130 nm gap) is shown B) A block diagram of the microscopy-based interrogation setup used to interrogate the photonic devices is shown. C) The 11 µm radius ring resonator used here shows a FSR of ≈ 9.2 nm near 1550 nm. D) The ring resonator acts as a notch filter whose resonance window is sensitive to temperature changes. The ring's resonance wavelength systematically increases as the temperature increases; resonances at various temperatures are shown in the insert**

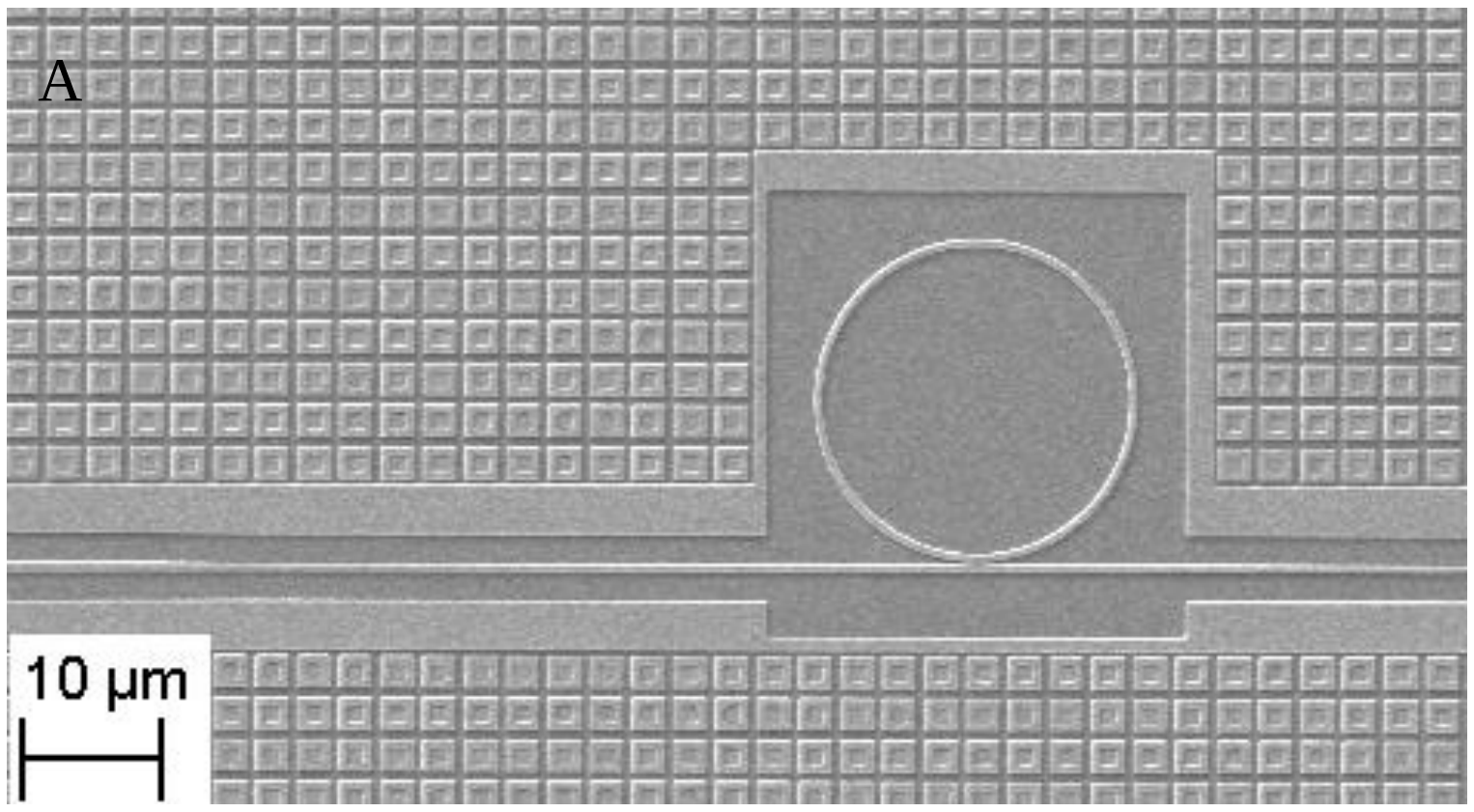

B

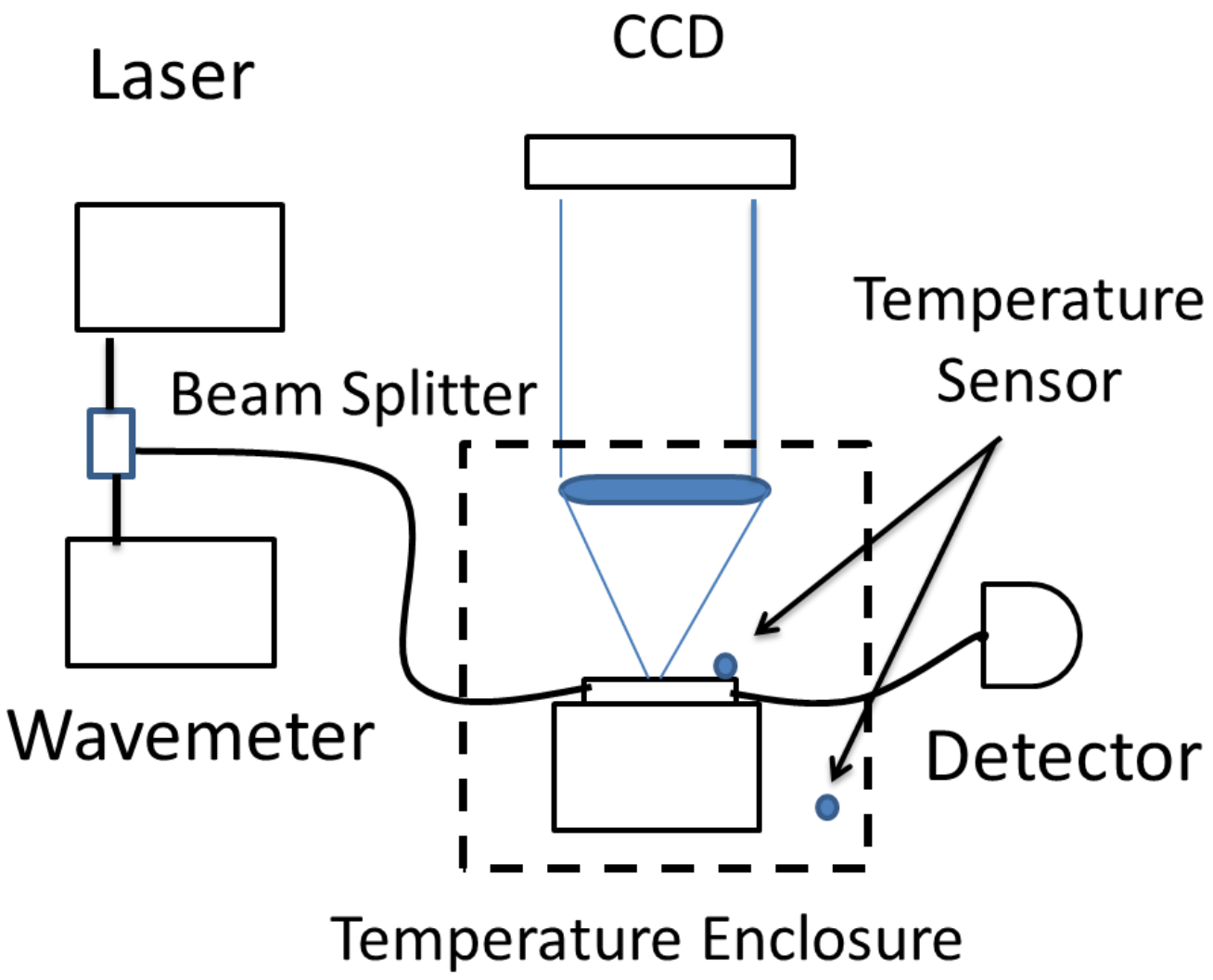

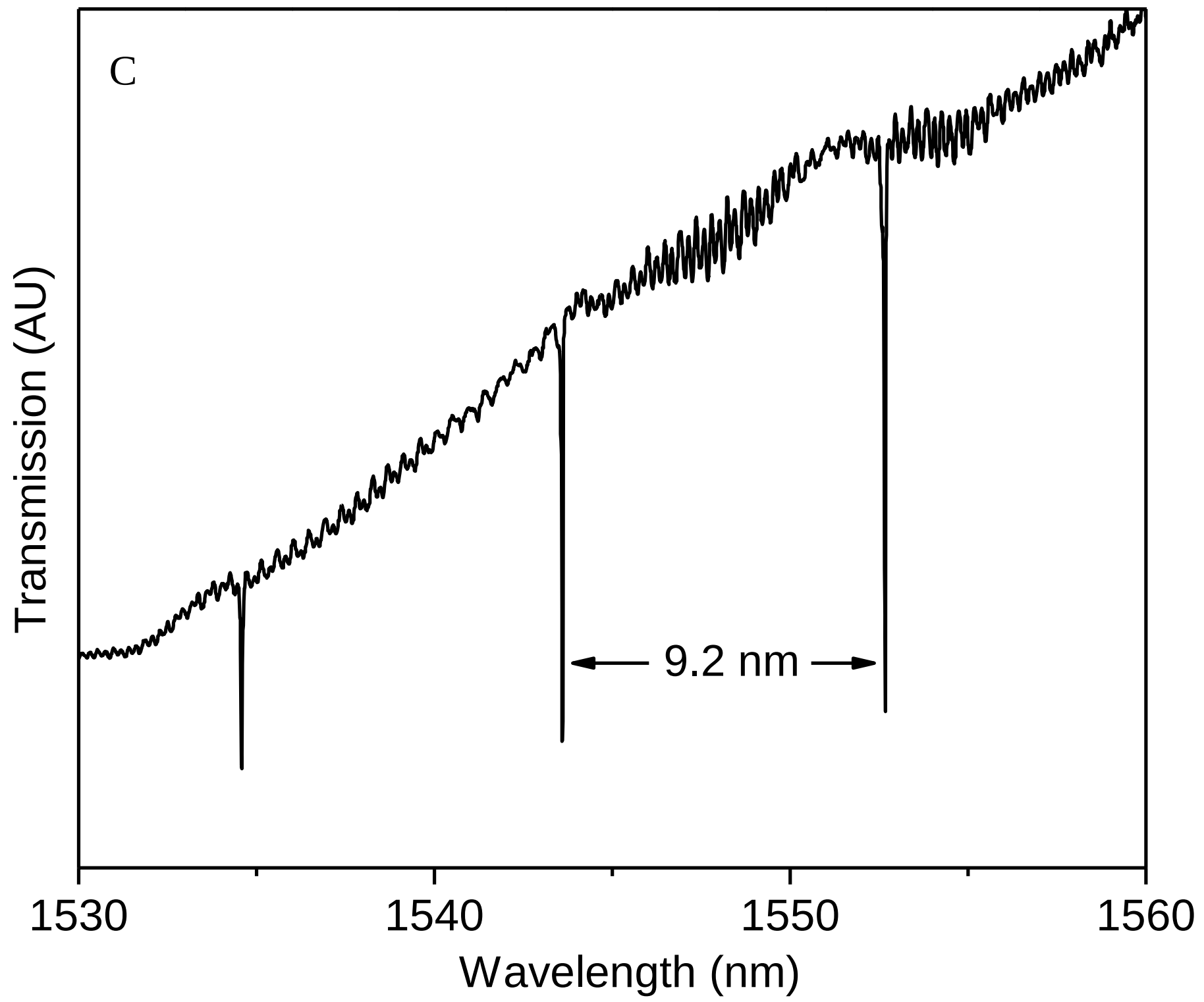

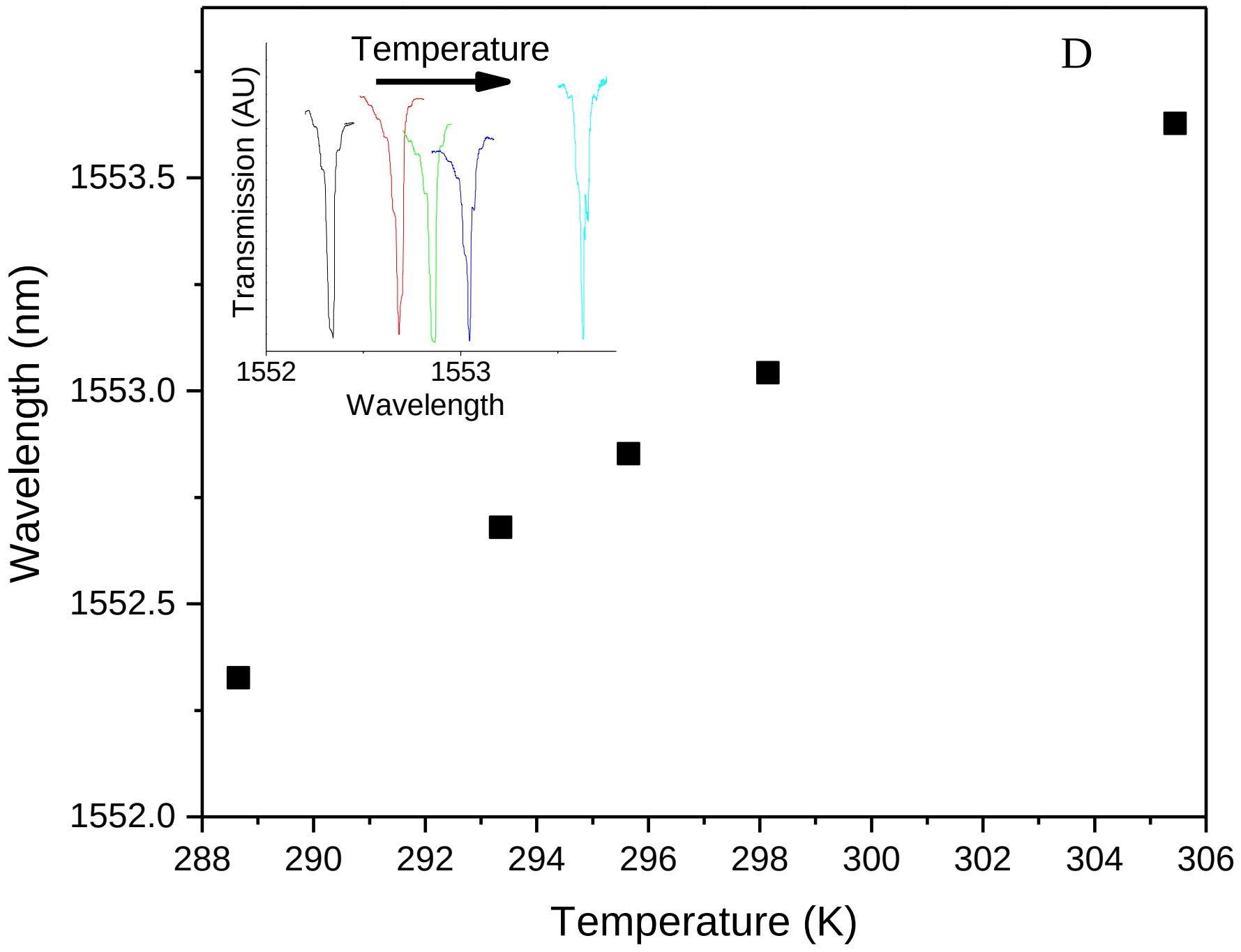

**Figure 2: a) Increasing incident laser power causes self-heating in ring resonator based devices resulting in an increase of the device resonance wavelength. Consequently, ring resonator increasingly over-estimates the ambient temperature compared to a platinum resistance thermometer. Over the incident power range of 0.0063 mW to 0.1 mW the estimated systematic temperature error is below 0.1 K. B) Power spectral density plot shows 1/f noise dependence. C) Allan Variance measurements indicate the instrument noise bottoms out at a 1 Hz measurement rate, creating a noise floor of ≈80 μK.**

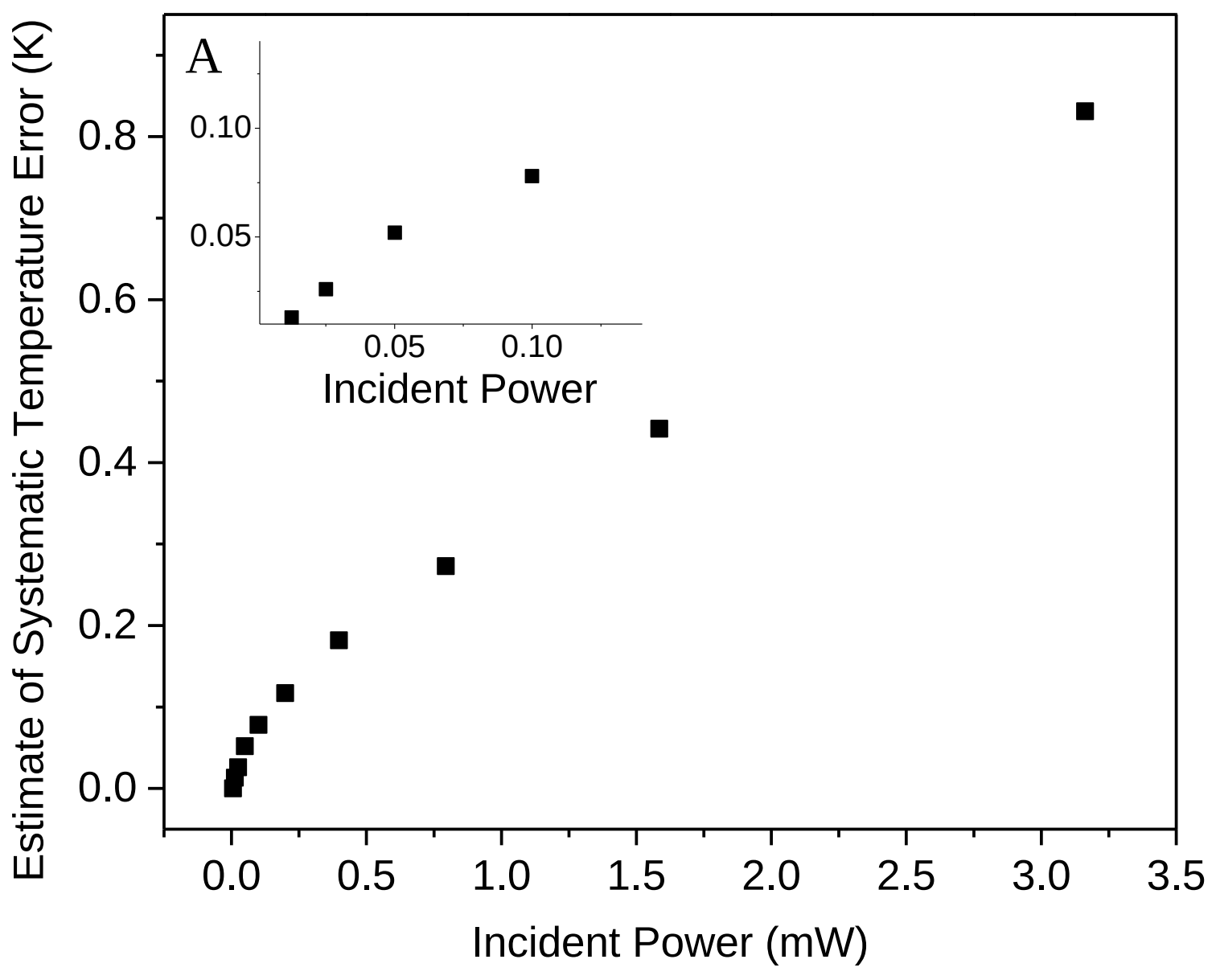

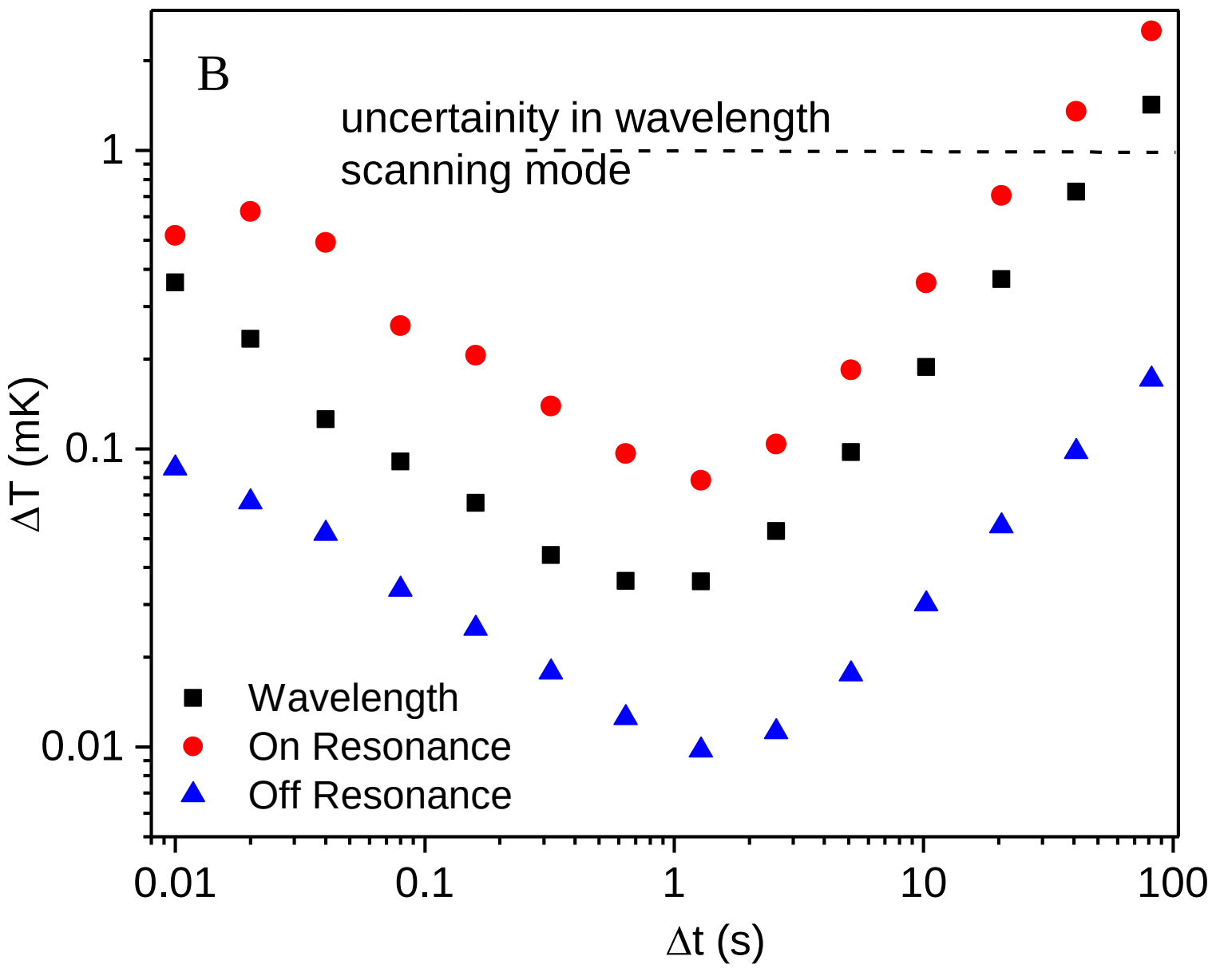

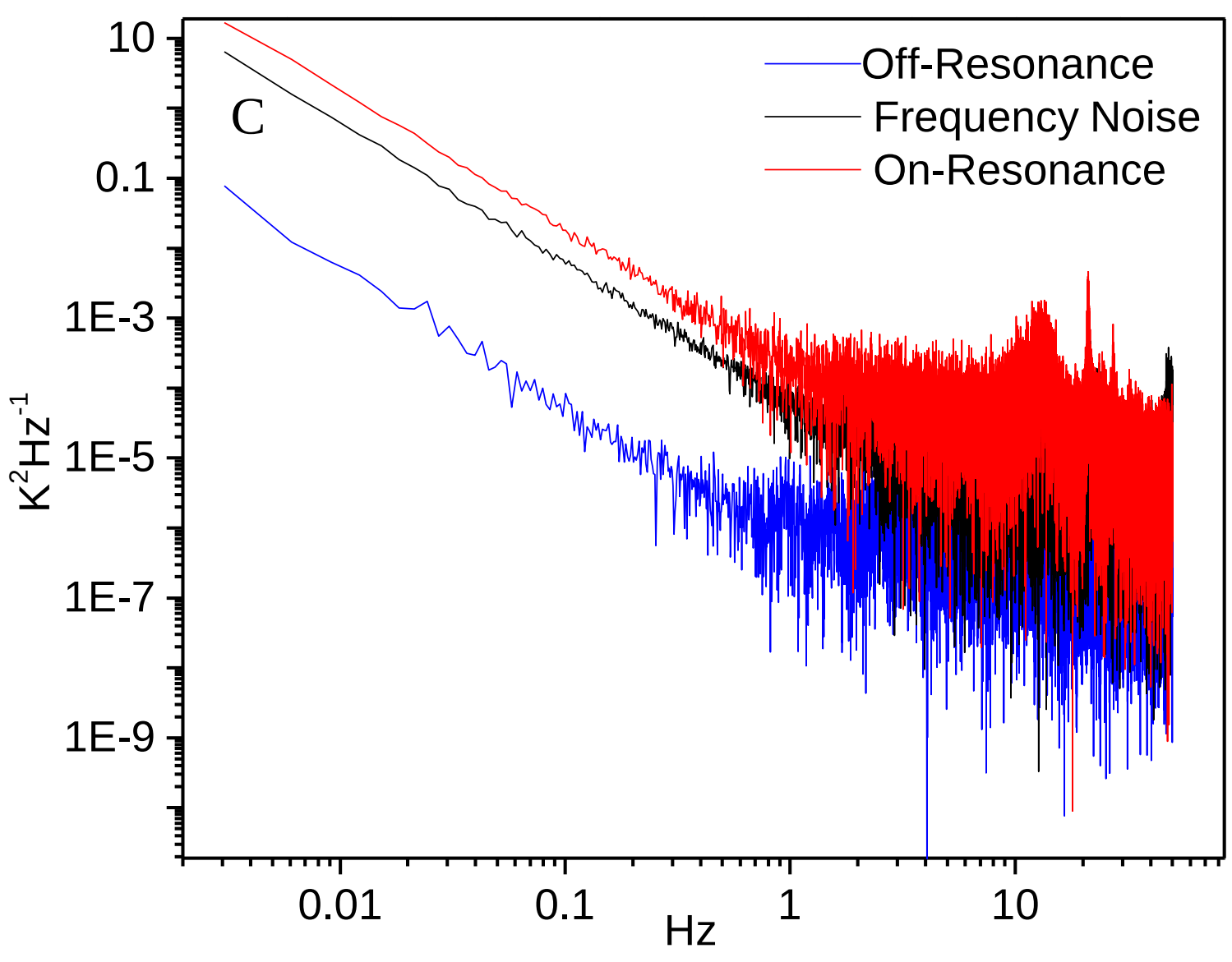